\documentclass[fleqn,usenatbib]{mnras} 

\usepackage{newtxtext,newtxmath}

\usepackage[T1]{fontenc}

\DeclareRobustCommand{\VAN}[3]{#2}
\let\VANthebibliography\thebibliography
\def\thebibliography{\DeclareRobustCommand{\VAN}[3]{##3}\VANthebibliography}

\usepackage{amsmath}
\usepackage{esint} 
\usepackage{siunitx}
\usepackage{lmodern}
\usepackage{graphicx}	

\title[Rotation Effects on Solar Granule]{How Dose Rotation Effect Convection Criterion and Solar Granule Size}

\author[Haibin Chen et al.]{
	Haibin Chen,
	Rong Wu\thanks{wurong2@mail3.sysu.edu.cn}
	\\
    School of Aeronautics and Astronautics, Sun Yat-sen University, Xin Gang Road W, Guangzhou 510275, China\\
}

\date{Accepted XXX. Received YYY; in original form ZZZ}

\pubyear{2023}

\begin{document}
	\label{firstpage}
	\pagerange{\pageref{firstpage}--\pageref{lastpage}}
	\maketitle
	
	\begin{abstract}

    The rotation of granules shows thermal properties similar to the thermal motion of molecules, which effects convection criterion and solar granule size.
    The granular rotation generates the rotational additional pressure and corresponding rotational equivalent temperature in the spherically symmetric expansion of the granule. 
    The rotation equivalent temperature represented by the rotational speed and the granular size is combined with the Schwarzschild criterion to obtain a new convection criterion determined jointly by the granular size, temperature gradient, and rotational speed gradient.
    In most of the solar convection zone, the convection is driven by the temperature gradient and suppressed by the rotational speed gradient.
    In the solar quiet region, the larger the granular size, the weaker the convection. 
    The small granules with the sizes smaller than the critical size are in natural convection, while the large granules with the sizes larger than the critical size are in forced convection.
    In the solar active region, the larger the rotational speed, the smaller the critical size and the temperature. 
    So sunspots have smaller granules and lower temperatures.

	\end{abstract}
	
	
	
	
	\section{Introduction}
	
	Solar granulation is the most obvious structures from the solar convection zone to the photosphere. 
	Observed granulation on the solar surface appear as a cellular pattern, which is formed by declining granules \citep{Oda1984}. 
	A large number of observations fit well into this scenario\citep{Han2016}: $\left( 1 \right)$ there is a critical granular diameter as the boundary for dividing granules into large granules and small granules; 
	$\left( 2 \right)$ The number of large granules decreases rapidly with the increase of diameter, while the number of small granules increases monotonously or remains flat with the decrease of diameter. 
    Due to different observation technologies and classification standards, the critical diameter has different values, such as 
    $1 ''. 37 \left( \approx 990 \mathrm{km} \right)$  \citep{Roudier1986}, 
    $1 ''. 4 \left( \approx 1000 \mathrm{km} \right)$  \citep{Hirzberger1997}, 
    $1000 \mathrm{km}$ \citep{Berrilli2002}, 
    $1 ''. 44$ \citep{Yu2011}, 
    and $600 \mathrm{km}$ \citep{Abramenko2012}.
    The above granulation phenomena are mainly aimed at the solar quiet regions, it is found that the granules in active regions manifest smaller sizes than those in quiet regions and are so-called abnormal granules, which is considered as the result of the solar magnetic field \citep{Title1992,Narayan2010}. 
	
	Granules are widely considered as evidences and the smallest cells of the solar interior convection, and the convection mechanism of granules is very important for the studies of solar quiet regions and active regions.
	There are many studies on the sizes and dynamic motions of the solar photospheric granules. 
	Granules are generally considered highly dynamic and turbulent, characteried by random motions and large Reynolds numbers.
	Roudier and Muller \citep{Roudier1986} suggested that small granules might be the turbulent origin, and the large granules are convective.
	Liu et al. \citep{Liu2021} find a critical size of $600 \mathrm{km}$ separating the granules in motion into two regime: the large granules dominated by convection and the small granules dominated by turbulence. The granules in active regions evolve with the absence of the mixing motions of convection and turbulence \citep{Liu2022}.
	Hirzberger et al. \citep{Hirzberger1999} hold that the smaller granules are more concentrated along downdraughts whereas larger ones preferentially occupy the updraughts.
    Kolmogorov \citep{Kolmogorov1991local,Kolmogorov1991dissipation} 
    thought that the cause of small granules increasing significantly with the decreasing sizes could be a cascade process, that is, considering vortices formed by shear instability, enery is tranported from large granules to small ones and at last dissipated by viscosity. 
    This similar turbulent cascade process has been mentioned in some literatures \citep{Salucci1994,Espagnet1995,Hirzberger1999}. 
    Kolmogorov theory is under the assumption of isotropic and homogeneous. 
	But Petrovay \citep{Petrovay2001} indicated that the strongly inhomogeneous and anisotropic conditions need to be considered to obtain a more complicated theory than Kolmogorov theory.  
	
	As such, the dynamic motions of granules might be a cascade process, but it lacks more details. 
	One of the most important points might be that people do not understand the specific energy conversion details of individual granules.
	In the study of solar differential rotation \citep{Chen2022}, the author found that the rotations of fluid cells will generate additional pressures under the assumption of spherically symmetric expansion. For a single fluid cell, the work done by the additional pressure leads to the change of rotational kinetic energy. Further, the additional pressures will effect the density and buoyancy of granules. At this time, the Schwarzschild criterion \citep{1958Structure,Thorne1966,Lin2000} will no longer be perfectly applicable, in other words, the convection criterion will not only be determined by the temperature gradient, but also by the rotational speed gradient.
	In this paper, we try to supplement more detailed  theoretical calculations to form a complete theory to help people gain insight into the convection mechanism of individual granules.

	\section{Thermal Properties of Rotating Solar Granule}
	
	\subsection{ Assumption of Spherical Symmetrical Expansion of Solar Granules}
	
	
	Due to the high Reynolds numbers in the solar convection zone, random motion that dominates the motion of solar granules is similar to the thermal motion of molecules. 
	Any individual granule could be simplified as a rotating "molecule" with a size and a irregular motion, and the "molecular" granule would still expand or be compressed as the external pressure changes. 
	This assumption abandons the  more complex thermal convection models and turbulence models, which is more advantageous to study some essential properties of rotating turbulent thermal convection. 
	
	The mean granular lifetime is about 5 to 16 minutes \citep{bahng1961lifetime,mehltretter1978balloon,alissandrakis1987determination,Hirzberger1999}, which is 4 or 5 orders of magnitude smaller than the solar rotation period $T_ 0$ of about 27 days \citep{thompson1996}.
	In the process of a granule expansion, there may be a rotational speed difference between it and the ambient fluid. 
	According to the propagation law of the inertial wave of the rotating fluid, the rotational speed difference of the granules will be transmitted to the outside by the inertial wave with a period of $\frac{T_0}{2}$.
	The granular lifetime is much shorter than the inertial wave period, which means that the granular rotational speed difference generated by the expansion is stuck in the granule itself.
	 
    The expansion direction can be considered independent from the rotation of a granule. 
	In the absence of other effects, the expansion of granulation is statistically close to isotropic and homogeneous. 
	For a individual granule, its expansion can be assumed to be spherically symmetric. 
	
	\subsection{Rotational Kinetic Energy and Rotational Additional Pressure}
	
	Ignoring viscosity, there is a rotational speed difference between a granule and its ambient fluid during granular spherically symmetric expansion. It is the obvious evidence of the rotational kinetic energy change.
	According to the conservation of angular momentum, the relationship between the rotational speed $\Omega$ and size $a$ of the granule is
	\begin{equation}\label{Eq.1}
		\Omega = \Omega_0 \left( \frac{a}{a_0} \right) ^{-2},
	\end{equation}
	where $\Omega_0$ and $a_0$ are the initial rotational speed and size of the granule. 
	The differential form of Eq.(\ref{Eq.1}) is
	\begin{equation}\label{Eq.2}
		\frac{\mathrm{d} \Omega}{\Omega} = -2 \frac{\mathrm{d} a}{a} ,
	\end{equation}
	where $\mathrm{d} \Omega$ and $\mathrm{d} a$ are the changes of the rotational speed and size of the granule respectively.
	
	In order to study the change of rotational kinetic energy, a cylindrical granule with a radius of $a$, a height of $2a$ and a rotational speed of $\Omega$ is taken as the model.
	The rotational kinetic energy $E_\Omega$ of the cylindrical granule is calculated as
	\begin{equation}\label{Eq.3}
		E_\Omega = \frac{1}{2} \pi \rho \Omega^2 a^5  , 
	\end{equation}
	or 
	\begin{equation}\label{Eq.4}
		E_\Omega = \frac{1}{4} M \Omega^2 a^2   ,
	\end{equation}
	where $\rho$ and $M$ are the density and mass of the granule.  
	The differential form of Eq.(\ref{Eq.3}) and Eq.(\ref{Eq.4}) is
	\begin{equation}\label{Eq.5}
		\frac{\mathrm{d} E_\Omega}{E_\Omega} = -2 \frac{\mathrm{d} a}{a} .
	\end{equation}
    where $\mathrm{d} E_\Omega$ is the change of the rotational kinetic energy of the granule.

    The granular spherically symmetric expansion can be considered as adiabatic because of no viscous dissipation.
	The change of granular rotational kinetic energy $\mathrm{d} E_\Omega$ requires the rotation to provide a additional pressure $p_\Omega$ to do equivalent work in the radial direction of the cylindrical granule.
	The work done by the rotational additional pressure is equal to
	\begin{equation}\label{Eq.6}
		\mathrm{d} E_\Omega = - \oiint p_\Omega \mathrm{d} l \mathrm{d} S .
	\end{equation}
	Calculate the average pressure on the granular surface for the rotational additional pressure, letting
	\begin{equation}\label{Eq.7}
		\bar{p}_\Omega = - \frac{\mathrm{d} E_\Omega}{\mathrm{d} V} .	
	\end{equation}
	According to $\frac{\mathrm{d} V }{V} = 3 \frac{\mathrm{d} a }{a}$ and $V = 2 \pi a^3$, by combining and solving Eq.(\ref{Eq.3}), Eq.(\ref{Eq.5}) and Eq.(\ref{Eq.7}), the average rotational additional pressure is obtained as
	\begin{equation}\label{Eq.8}
		\bar{p}_\Omega = \frac{1}{6} \rho \Omega^2 a^2.
	\end{equation}
    It can be seen that, the larger the granule size, the greater the additional pressure.
    
	Similarly, for different shapes of granules, the average rotational additional pressure and rotational kinetic energy can be expressed as
	\begin{equation}\label{Eq.9}
		\bar{p}_\Omega = k_\Omega \rho \Omega^2 a^2
	\end{equation}
    and 
    \begin{equation}\label{Eq.10}
    	E_\Omega = \frac{3}{2} k_\Omega M \Omega^2 a^2
    \end{equation}
    respectively, where $k_\Omega$ is a constant related to the shape of the granule.  $k_\Omega$ has different values for different shapes.
	
	Since the solar rotation period is several orders of magnitude longer than the granular lifetime, the rotational speed difference between the granule and the ambient fluid is allowed to exist. 
	This means that when we study the equilibrium state of the granule, we should refer to the equilibrium of pressure, rather than the equilibrium of rotational speed. 
	There may be some part of pressure imbalances when the overall pressure of the granule is in equilibrium.
	That is, compared with $p_\Omega$ , $\bar{p}_\Omega$  is more suitable for describing the rotational additional pressure generated by the granular rotation and analysing the equilibrium state of the granule.
	
	\subsection{Hypothesis of rotational equivalent temperature }
	
	The granular rotation not only generates the rotational additional pressure on the granular surface, but also increases the overall granular kinetic energy. 
	This is similar to the relationship among temperature, pressure and molecular kinetic energy in molecular thermal motion. 
	As such, it is probably easier to illustrate the effects of the granular rotation by analogy with molecular thermal motion.
	
	According to ideal gas law, the pressure provided by molecular thermal motion is $p = \frac{R_\mathrm{m}}{M_\mathrm{m}} \rho T$ , where $R_\mathrm{m}$ is the molar gas constant and $M_\mathrm{m}$ is the molar mass of the gas. 
	The rotational additional pressure can also be expressed in this form, that is, 
	\begin{equation}\label{Eq.11}
		\bar{p}_\Omega = \frac{R_\mathrm{m}}{M_\mathrm{m}} \rho T_\Omega   ,
	\end{equation}
	where $T_\Omega$ is an equivalent temperature corresponding to the rotational additional pressure in Eq.(\ref{Eq.9}), and rotational equivalent temperature $T_\Omega $ can be calculated as
	\begin{equation}\label{Eq.12}
		T_\Omega =\frac{M_\mathrm{m}}{R_\mathrm{m}} k_\Omega \Omega^2 a^2   .
	\end{equation}
    The differential form of Eq.(\ref{Eq.12}) is
    \begin{equation}\label{Eq.13}
    	\frac{\mathrm{d} T_\Omega}{T_\Omega} = -2 \frac{\mathrm{d} a}{a} .
    \end{equation}
    where $\mathrm{d} T_\Omega$ is the change of the rotational equivalent temperature of the granule. 
	
	The relationship between rotational additional pressure and rotational kinetic energy has been shown in Eq.(\ref{Eq.9}) and Eq.(\ref{Eq.10}).
	Futher study on the relationship between rotational equivalent temperature and rotational kinetic energy is necessary. 
	
	For a rotating granule, by combining and deriving Eq.(\ref{Eq.5}), Eq.(\ref{Eq.10}), Eq.(\ref{Eq.12}) and Eq.(\ref{Eq.13}), we get the change of rotational kinetic energy $\mathrm{d} E_\Omega$ caused by the change of rotational equivalent temperature $\mathrm{d} T_\Omega$, that is 
	\begin{equation}\label{Eq.14}
		\mathrm{d} E_\Omega = \frac{3}{2} \frac{M}{M_\mathrm{m}} R_\mathrm{m} \mathrm{d} T_\Omega   . 
	\end{equation}
    Obviously, Eq.(\ref{Eq.14}) is similar to the formula for calculating the internal energy change of an ideal gas. 
	Comparing with the definition of molar specific heat capacity at constant volume in thermodynamics, it can be considered that the molar equivalent specific kinetic energy capacity at constant volume of the rotating cylindrical granule is
	\begin{equation}\label{Eq.15}
	    C_{V_\Omega} = \frac{\mathrm{d} E_\Omega}{\frac{M}{M_\mathrm{m}} \mathrm{d} T_\Omega} = \frac{3}{2} R_\mathrm{m}  , 
	\end{equation}
	which is the same as the molar specific heat capacity at constant volume of a monatomic ideal gas. 	
		  
	These conducted results indicate that the rotation of a granule will lead to some equivalent thermal properties similar to molecular thermal motion, such as $T_\Omega$, $p_\Omega$, $E_\Omega$ and $C_{V_\Omega}$. 	
	In the spherically symmetric expansion of a rotating granule, $T$ and $T_\Omega$ are coupled. 
	The temperature-related process is the adiabatic expansion of an ideal gas with intramolecular degrees of freedom $i$ and a specific heat ratio $\gamma$.
	The rotation-related process can be considered as the equivalent adiabatic expansion of an monatomic ideal gas with intramolecular degrees of freedom $i_\Omega = 3$ and a specific heat ratio $\gamma_\Omega = \frac{5}{3}$. 

	\section{Convection Criterion Effected by Rotation}
	
	\subsection{Derivation of convection criterion effected by rotation}
	
	The rotational additional pressure and its corresponding rotational equivalent temperature can effect the density and buoyancy of the granule. Further more, the rotatioanl additional pressure changes the equilibrium state of the granule and the rotational equivalent temperature acts on the convection criterion like the temperature.                                                               
	
	The Schwarzschild criterion for irrotational granule is
	\begin{equation}\label{Eq.16}
	    \left| \frac{\mathrm{d} T}{\mathrm{d} l} \right| _{\mathrm{rd}} > \left| \frac{\mathrm{d} T}{\mathrm{d} l} \right| _{\mathrm{ad}} ,
	\end{equation}
	where $\left( \frac{\mathrm{d} T}{\mathrm{d} l} \right) _{\mathrm{rd}}$ is the real temperature gradient of the granule, $\left( \frac{\mathrm{d} T}{\mathrm{d} l} \right) _{\mathrm{ad}}$ is the adiabatic temperature gradient and can be expressed as
	\begin{equation}\label{Eq.17}
        \left( \frac{\mathrm{d} T}{\mathrm{d} l} \right) _{\mathrm{ad}} = \left( 1-\frac{1}{\gamma} \right) T \frac{\mathrm{d} p}{p \mathrm{d} l} ,
	\end{equation}
	or
	\begin{equation}\label{Eq.18}
    	\left( \frac{\mathrm{d} T}{\mathrm{d} l} \right) _{\mathrm{ad}} = \left( \gamma - 1 \right) T \frac{\mathrm{d} \rho}{\rho \mathrm{d} l} .
	\end{equation}
    For any ideal gas, there is $\gamma > 1$ and $\frac{\mathrm{d} \rho}{\rho \mathrm{d} l}$ is always positive or negative. The Schwarzschild criterion can be described as: 
    \begin{equation}\label{Eq.19}
    	\left\{
    	\begin{matrix} 
    		\text{if} \, \frac{\mathrm{d} \rho}{\rho \mathrm{d} l} < 0,
    		\text{then} \, \frac{\mathrm{d} T}{\mathrm{d} l} < \left( \gamma - 1 \right) T \frac{\mathrm{d} \rho}{\rho \mathrm{d} l};  
    		\\  
    		\text{if} \, \frac{\mathrm{d} \rho}{\rho \mathrm{d} l} > 0, 
    		\text{then} \, \frac{\mathrm{d} T}{\mathrm{d} l} > \left( \gamma - 1 \right) T \frac{\mathrm{d} \rho}{\rho \mathrm{d} l} . 
    	\end{matrix}\right. 
    \end{equation}
    
	Considering the effect of the rotational equivalent temperature, criterion of rotating turbulent thermal convection becomes
	\begin{equation}\label{Eq.20}
    	\left| \frac{\mathrm{d} T  \, \text{+} \, \mathrm{d} T_\Omega}{\mathrm{d} l} \right| _{\mathrm{rd}} > \left| \frac{\mathrm{d} T  \, \text{+} \,  \mathrm{d} T_\Omega}{\mathrm{d} l} \right| _{\mathrm{ad}} ,
	\end{equation}
	where $\left( \frac{\mathrm{d} T_\Omega}{\mathrm{d} l} \right) _{\mathrm{rd}}$ is the real rotational equivalent temperature gradient of the granule, and $\left( \frac{\mathrm{d} T_\Omega}{\mathrm{d} l} \right) _{\mathrm{ad}}$ is the adiabatic rotational equivalent temperature gradient.
	Imitating Eq.(\ref{Eq.18}), one has 
	\begin{equation}\label{Eq.21}
    	\left( \frac{\mathrm{d} T_\Omega}{\mathrm{d} l} \right) _{\mathrm{ad}} = \left( \gamma_\Omega - 1 \right) T_\Omega \frac{\mathrm{d} \rho}{\rho \mathrm{d} l} = \frac{2}{3} T_\Omega \frac{\mathrm{d} \rho}{\rho \mathrm{d} l} .
	\end{equation}

	By taking the derivative of Eq.(\ref{Eq.11}) with respect to $l$, we get
	\begin{equation}\label{Eq.22}
    	\left( \frac{\mathrm{d} T_\Omega}{\mathrm{d} l} \right) _{\mathrm{rd}}  = T_\Omega \left( 2 \frac{\mathrm{d} \Omega }{\Omega \mathrm{d} l} \, \text{+} \, 2 \frac{\mathrm{d} a }{a \mathrm{d} l} \right) ,
	\end{equation}
	and 
	\begin{equation}\label{Eq.23}
    	\left( \frac{\mathrm{d} T_\Omega}{\mathrm{d} l} \right) _{\mathrm{ad}}  = T_\Omega \left[ 2 \left( \frac{\mathrm{d} \Omega }{\Omega \mathrm{d} l} \right) _{\mathrm{ad}} \, \text{+} \, 2 \frac{\mathrm{d} a }{a \mathrm{d} l} \right] .
	\end{equation}
    For a single granule, according to the law of mass conservation, the relationship between the size and density of the granule is
	\begin{equation}\label{Eq.24}
    	\frac{\mathrm{d} a }{a \mathrm{d} l} = - \frac{1}{3} \frac{\mathrm{d} \rho }{\rho \mathrm{d} l}   .
	\end{equation}
    By substituting Eq.(\ref{Eq.24}) into Eq.(\ref{Eq.22}), we get
    \begin{equation}\label{Eq.25}
    	\left( \frac{\mathrm{d} T_\Omega}{\mathrm{d} l} \right) _{\mathrm{rd}}  = T_\Omega \left( 2 \frac{\mathrm{d} \Omega }{\Omega \mathrm{d} l} - \frac{2}{3} \frac{\mathrm{d} \rho }{\rho \mathrm{d} l} \right) .
    \end{equation}
	Combining and solving Eq.(\ref{Eq.21}), Eq.(\ref{Eq.23}) and Eq.(\ref{Eq.24}), then we get a new expression of adiabatic gradient, that is
	\begin{equation}\label{Eq.26}
    	\left( \frac{\mathrm{d} \Omega}{\Omega \mathrm{d} l} \right) _\mathrm{ad} = \frac{2}{3} \frac{\mathrm{d} \rho }{\rho \mathrm{d} l} .
	\end{equation}
	This means that we might be able to use rotational speed $\Omega$ instead of rotational equivalent temperature $T_\Omega$ to directly describe the effect of granular rotation on convection criterion.  
	
	Substituting Eq.(\ref{Eq.18}), Eq.(\ref{Eq.21}) and Eq.(\ref{Eq.25}) into Eq.(\ref{Eq.20}), the new convection criterion in Eq.(\ref{Eq.19}) is expressed as
	\begin{equation} \label{Eq.27}
			\left| \frac{\mathrm{d} T }{\mathrm{d} l} \, \text{+} \, T_\Omega \left( 2 \frac{\mathrm{d} \Omega }{\Omega \mathrm{d} l} -\frac{2}{3} \frac{\mathrm{d} \rho}{\rho \mathrm{d} l} \right) \right|  >
			\left|  \left( \gamma - 1 \right) T \frac{\mathrm{d} \rho}{\rho \mathrm{d} l}  \, \text{+} \, \frac{2}{3} T_\Omega \frac{\mathrm{d} \rho}{\rho \mathrm{d} l}  \right|  .
	\end{equation}
    
	Further, substituting Eq.(\ref{Eq.12}) into Eq.(\ref{Eq.27}), the convection criterion in Eq.(\ref{Eq.27}) becomes
	\begin{equation} \label{Eq.28}
    	\begin{split}	
    	&\left| \frac{\mathrm{d} T }{\mathrm{d} l} \, \text{+} \,  \frac{M_\mathrm{m}}{R_\mathrm{m}} k_\Omega \Omega^2 a^2 \left(2 \frac{\mathrm{d} \Omega }{\Omega \mathrm{d} l} -\frac{2}{3} \frac{\mathrm{d} \rho}{\rho \mathrm{d} l} \right) \right|  >
    	\\ &\left|  \left( \gamma - 1 \right) T \frac{\mathrm{d} \rho}{\rho \mathrm{d} l}  \, \text{+} \, \frac{2}{3}  \frac{M_\mathrm{m}}{R_\mathrm{m}} k_\Omega \Omega^2 a^2 \frac{\mathrm{d} \rho}{\rho \mathrm{d} l}  \right|  .
    	\end{split}
	\end{equation}
    Eq.(\ref{Eq.28}) can be considered as the general forrmula of the new convection criterion effected by rotation. Comparing with Eq.(\ref{Eq.16}), it can be seen that the convection criterion will be not only determined by the temperature gradient, but also effected by the rotational speed gradient and the granule size.  
    After rearranging, the new convection criterion in Eq.(\ref{Eq.28}) can be described as: 
    \begin{equation}\label{Eq.29}
    	\left\{
    	\begin{matrix} 
    		\begin{split}
    			&\text{if} \, \frac{\mathrm{d} \rho}{\rho \mathrm{d} l} < 0, 
    			\text{then} \\ 
    			&2 \frac{M_\mathrm{m}}{R_\mathrm{m}} k_\Omega \Omega^2 a^2 \left( \frac{\mathrm{d} \Omega }{\Omega \mathrm{d} l} -\frac{2}{3} \frac{\mathrm{d} \rho}{\rho \mathrm{d} l} \right) 
    			< 
    			- T \left[ \frac{\mathrm{d} T }{T \mathrm{d} l} - \left( \gamma - 1 \right) \frac{\mathrm{d} \rho}{\rho \mathrm{d} l} \right]  ; 
    		\end{split} 
    		\\ 
    		\begin{split} 
    			&\text{if} \, \frac{\mathrm{d} \rho}{\rho \mathrm{d} l} > 0, 
    			\text{then} \\ 
    			&2 \frac{M_\mathrm{m}}{R_\mathrm{m}} k_\Omega \Omega^2 a^2 \left( \frac{\mathrm{d} \Omega }{\Omega \mathrm{d} l} -\frac{2}{3} \frac{\mathrm{d} \rho}{\rho \mathrm{d} l} \right) 
    			> 
    			- T \left[ \frac{\mathrm{d} T }{T \mathrm{d} l} - \left( \gamma - 1 \right) \frac{\mathrm{d} \rho}{\rho \mathrm{d} l} \right] . 
    		\end{split} 
    	\end{matrix}\right. 
    \end{equation}
                                  
	\subsection{Analysis of convection criterion effected by rotation}
	
\begin{figure}
		\includegraphics[width=\columnwidth]{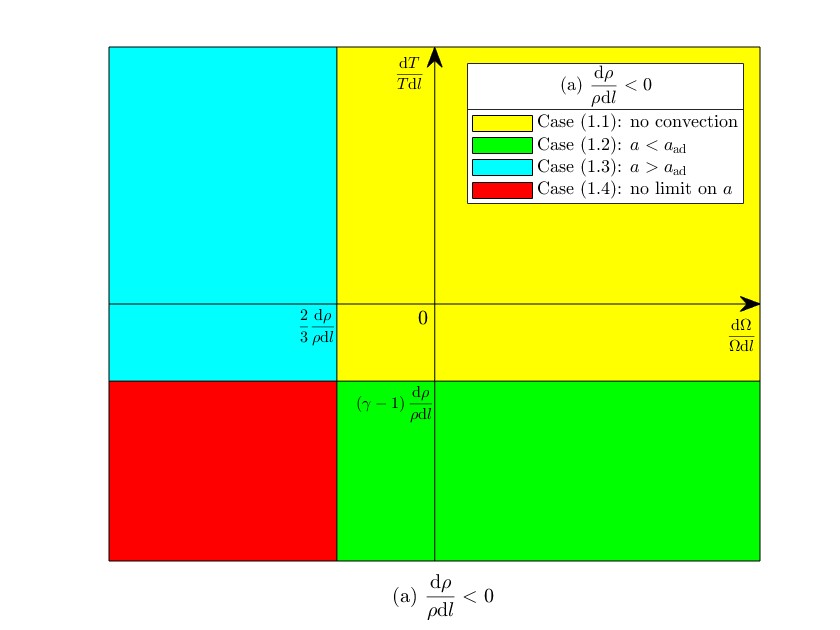}
		\includegraphics[width=\columnwidth]{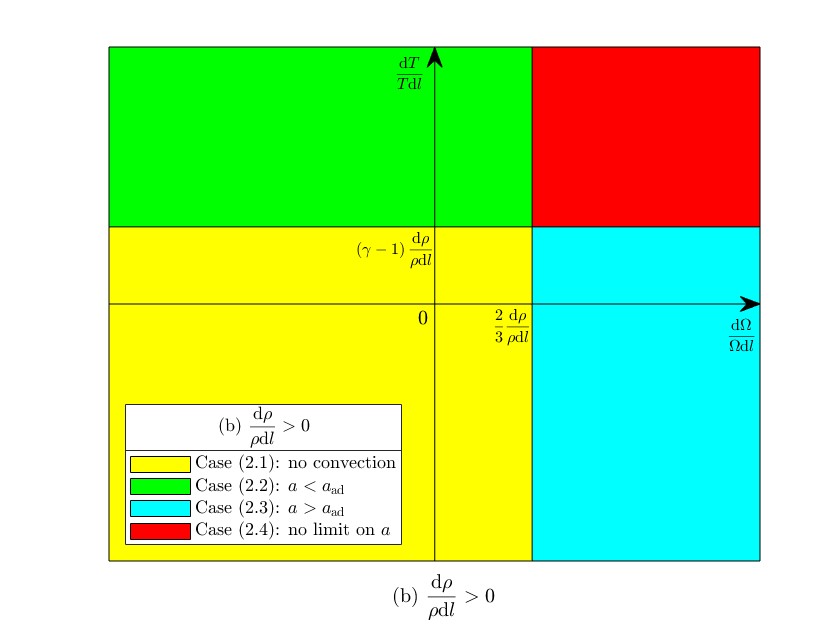}
		\caption{New convection criterion effected by rotation.}
		\label{fig: new criterion}
	\end{figure}

	For granules in different areas, the values of $T$, $\Omega$, $\frac{\mathrm{d} T}{\mathrm{d} l}$, $\frac{\mathrm{d} \Omega }{\Omega \mathrm{d} l}$ and $\frac{\mathrm{d} \rho}{\rho \mathrm{d} l}$ are different.
	For the individual granules in a certain area, the values of $a$ are not the same. 
	The critical condition for the inequality of the new convection criterion taking the equal sign is that the granule size $a$ takes the critical value recorded as $a_\mathrm{ad}$.
	The calculated critical size $a_\mathrm{ad}$ is
	\begin{equation}\label{Eq.30}
		a^2_\mathrm{ad} = \frac
		{- T \left[ \frac{\mathrm{d} T}{T \mathrm{d} l} - \left( \gamma - 1 \right) \frac{\mathrm{d} \rho}{\rho \mathrm{d} l} \right]}
		{\frac{2 k_\Omega M_m}{ R_m} \Omega^2 \left( \frac{\mathrm{d} \Omega}{\Omega \mathrm{d} l} - \frac{2}{3} \frac{\mathrm{d} \rho}{\rho \mathrm{d} l} \right)} .
	\end{equation}
	From the three factors $\frac{\mathrm{d} T}{T \mathrm{d} l}$, $\frac{\mathrm{d} \Omega}{\Omega \mathrm{d} l}$ and $a$, the establishment conditions of the new convection criterion in Eq.(\ref{Eq.29}) can be analysed and the results are as follows:
	\begin{itemize}
		\item (1) If $\frac{\mathrm{d} \rho}{\rho \mathrm{d} l} < 0$,
		\subitem
		Case (1.1): when $\frac{\mathrm{d} T}{T \mathrm{d} l} > \left( \gamma - 1 \right) \frac{\mathrm{d} \rho}{\rho \mathrm{d} l}$ and $\frac{\mathrm{d} \Omega}{\Omega \mathrm{d} l} > \frac{2}{3} \frac{\mathrm{d} \rho}{\rho \mathrm{d} l}$, the new convection criterion does not hold. This means that natural convection cannot occur without the temperature gradient or the rotational speed gradient to drive.
		\subitem 
		Case (1.2): when $\frac{\mathrm{d} T}{T \mathrm{d} l} < \left( \gamma - 1 \right) \frac{\mathrm{d} \rho}{\rho \mathrm{d} l}$ and $\frac{\mathrm{d} \Omega}{\Omega \mathrm{d} l} > \frac{2}{3} \frac{\mathrm{d} \rho}{\rho \mathrm{d} l}$, the new convection criterion becomes $a < a_\mathrm{ad}$. This means that the granules with sizes smaller than $a_\mathrm{ad}$ are in the convection driven by the temperature gradient and suppressed by the rotational speed gradient. 
		\subitem
		Case (1.3): when $\frac{\mathrm{d} T}{T \mathrm{d} l} > \left( \gamma - 1 \right) \frac{\mathrm{d} \rho}{\rho \mathrm{d} l}$ and $\frac{\mathrm{d} \Omega}{\Omega \mathrm{d} l} < \frac{2}{3} \frac{\mathrm{d} \rho}{\rho \mathrm{d} l}$, the new convection criterion becomes $a > a_\mathrm{ad}$. This means that the granules with sizes larger than $a_\mathrm{ad}$ are in the convection driven by the rotational speed gradient and suppressed by the temperature gradient.  
		\subitem
		Case (1.4): when $\frac{\mathrm{d} T}{T \mathrm{d} l} < \left( \gamma - 1 \right) \frac{\mathrm{d} \rho}{\rho \mathrm{d} l}$ and $\frac{\mathrm{d} \Omega}{\Omega \mathrm{d} l} < \frac{2}{3} \frac{\mathrm{d} \rho}{\rho \mathrm{d} l}$, the new convection criterion holds naturally. This means that the granules with any sizes are in the convection driven by both the temperature gradient and the rotational speed gradient.
		\item (2) If $\frac{\mathrm{d} \rho}{\rho \mathrm{d} l} > 0$,
		\subitem
		Case (2.1): when $\frac{\mathrm{d} T}{T \mathrm{d} l} < \left( \gamma - 1 \right) \frac{\mathrm{d} \rho}{\rho \mathrm{d} l}$ and $\frac{\mathrm{d} \Omega}{\Omega \mathrm{d} l} < \frac{2}{3} \frac{\mathrm{d} \rho}{\rho \mathrm{d} l}$, the new convection criterion does not hold. This means that natural convection cannot occur without the temperature gradient or the rotational speed gradient to drive.
		\subitem 
		Case (2.2): when $\frac{\mathrm{d} T}{T \mathrm{d} l} > \left( \gamma - 1 \right) \frac{\mathrm{d} \rho}{\rho \mathrm{d} l}$ and $\frac{\mathrm{d} \Omega}{\Omega \mathrm{d} l} < \frac{2}{3} \frac{\mathrm{d} \rho}{\rho \mathrm{d} l}$, the new convection criterion becomes $a < a_\mathrm{ad}$. This means that the granules with sizes smaller than $a_\mathrm{ad}$ are in the convection driven by the temperature gradient and suppressed by the rotational speed gradient. 
		\subitem
		Case (2.3): when $\frac{\mathrm{d} T}{T \mathrm{d} l} < \left( \gamma - 1 \right) \frac{\mathrm{d} \rho}{\rho \mathrm{d} l}$ and $\frac{\mathrm{d} \Omega}{\Omega \mathrm{d} l} > \frac{2}{3} \frac{\mathrm{d} \rho}{\rho \mathrm{d} l}$, the new convection criterion becomes $a > a_\mathrm{ad}$. This means that the granules with sizes larger than $a_\mathrm{ad}$ are in the convection driven by the rotational speed gradient and suppressed by the temperature gradient.  
		\subitem
		Case (2.4): when $\frac{\mathrm{d} T}{T \mathrm{d} l} > \left( \gamma - 1 \right) \frac{\mathrm{d} \rho}{\rho \mathrm{d} l}$ and $\frac{\mathrm{d} \Omega}{\Omega \mathrm{d} l} > \frac{2}{3} \frac{\mathrm{d} \rho}{\rho \mathrm{d} l}$, the new convection criterion holds naturally. This means that the granules with any sizes are in the convection driven by both the temperature gradient and the rotational speed gradient.
	\end{itemize}
	The results of the above complete discussions on the convection criterion effected by rotation can be shown in Fig.~\ref{fig: new criterion}. 
	
	\section{Solar Granule Size Effected By Rotation}
	
	\subsection{Solar granule size effected by rotation in the solar quiet region}
	
\begin{figure}
    	\includegraphics[width=\columnwidth]{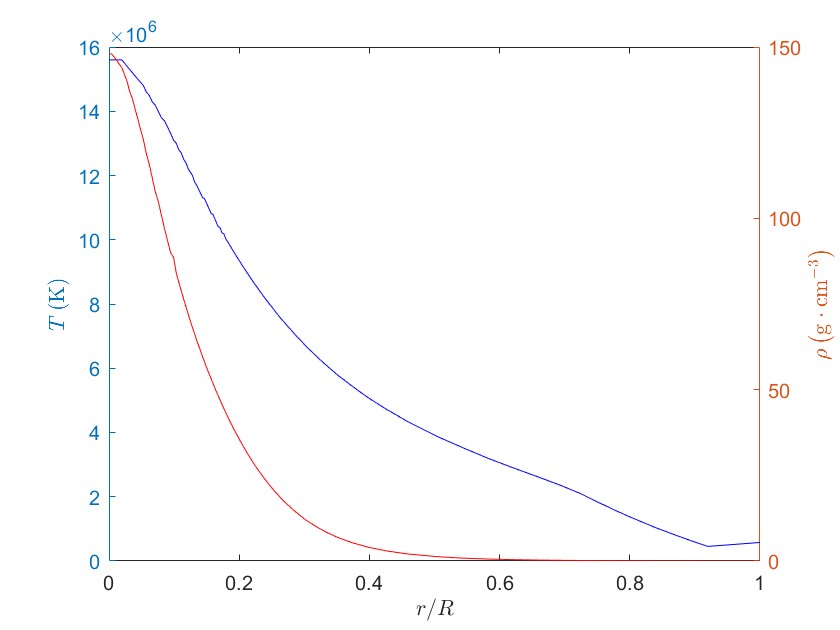}
    	\caption{Temperature and density variation with solar radius for a standard solar model.}
    	\label{fig: T and rho}
    \end{figure}

\begin{figure}
    \includegraphics[width=\columnwidth]{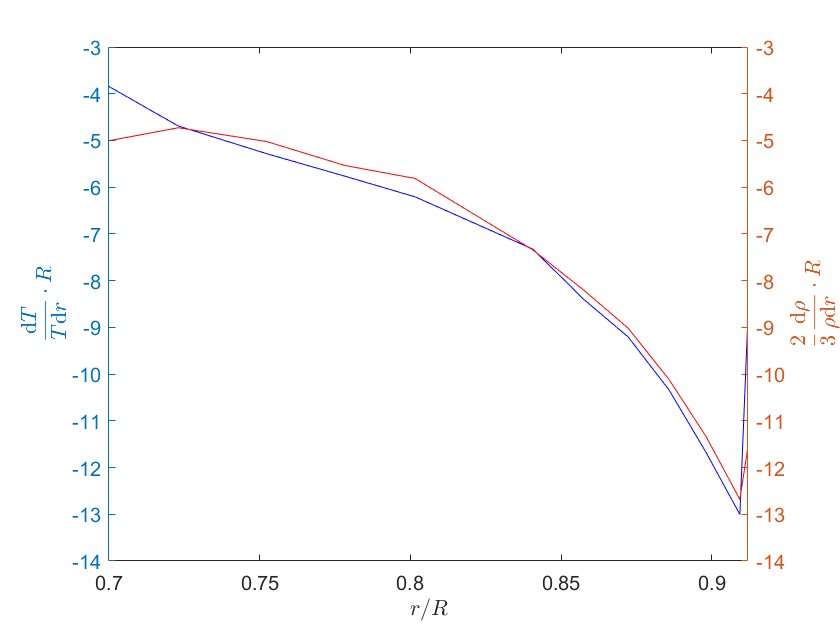}
    \caption{Relative temperature gradient and relative density gradient for a standard solar model.}
    \label{fig: gradients of T and rho}
    \end{figure}

	The models of the solar convection zone are mostly based on Schwarzschild criterion.
    \citet{christensen1991depth} conclude that the base of the adiabatically stratified region of the solar convection zone is at a radius $r_\mathrm{b} = 0.713\pm 0.003 R$, where $R_\odot$ is the solar radius; thus the depth of the solar convection zone is (0.287 ± 0.003) solar radii.
    \citet{Yang2015} introduce a commonly used standard solar model for extrapolating the solar internal structure. 
    Specific calculations for a standard solar model show that the region from about $0.75 R$ to the vicinity of the solar surface (bottom of the photosphere) satisfies Schwarzschild criterion and forms convection. 
    They give a set of data on the temperature and density inside the sun changing with the radius, and we could get the distributions of temperature, density and related gradients as shown in Fig.~\ref{fig: T and rho} and Fig.~\ref{fig: gradients of T and rho}. It can be seen that when $\gamma = \frac{5}{3}$, then $\frac{\mathrm{d} T}{T \mathrm{d} l} < \left( \gamma - 1 \right) \frac{\mathrm{d} \rho}{\rho \mathrm{d} l}$ in most solar convection zone.
    
	Based on the Schwarzschild criterion, we can also establish a solar convection zone model, where the convection is not only driven by temperature gradient but may also be driven or suppressed by rotational speed gradient.
	That is, the convection criterion for granules in the solar convection zone belongs to either Case(1.2) or Case(1.4) in section 3.2.
	
\begin{figure}
		\includegraphics[width=\columnwidth]{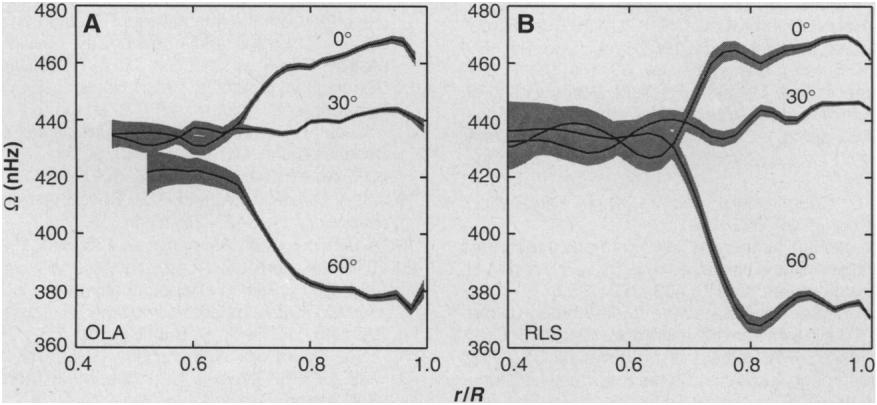}
		\caption{Solor differential rotation models based on GONG data by OLA and RLS.}
		\label{fig: differential rotation}
	\end{figure}

	The base of the solar convection zone is a region of transition not just for the temperature gradient, but also for the solar internal rotation \citep{basu1997seismology}.
	Some attempts to measure the solar internal rotational speed mainly relies on the observation data in the photospheric and chromospheric layers and the helioseismic inversion in the solar convection zone \citep{schroter1985}.
	For example, Fig.~\ref{fig: differential rotation} \citep{thompson1996} shows two solar differential rotation models which are inversions of GONG data by optimally  localized averages (OLA) and regularized least-squares (RLS) methods.
	Although there are differences in the solar differential rotation models in literatures, there is a consensus that the radial rotational speed gradient in the convection zone is generally small from helioseismology .
	
	When $\gamma = \frac{5}{3}$, the temperature gradient $\frac{\mathrm{d} T}{T \mathrm{d} l}$ and relative rotational speed gradient $\frac{\mathrm{d} \Omega}{\Omega \mathrm{d} l}$ have the same adiabatic value $\frac{2}{3} \frac{\mathrm{d} \rho}{\rho \mathrm{d} l}$, so it can be considered that the rotational speed gradient in the convection zone satisfying the Schwarzschild criterion does not reach the adiabatic value, i.e. $\frac{\mathrm{d} \Omega}{\Omega \mathrm{d} l} > \frac{2}{3} \frac{\mathrm{d} \rho}{\rho \mathrm{d} l}$.
	That is, the convection criterion for granules in most of the solar convection zone belongs to Case(1.2).
	
	In Case(1.2), convection of granules is driven by the temperature gradient and suppressed by the rotational speed gradient. 
	For a individual granule, it can be seen from Eq.(\ref{Eq.9}) and Eq.(\ref{Eq.12}) that the larger the granule size, the stronger the suppression effect of the rotation.
	If $a < a_\mathrm{ad}$, the driving effect of temperature gradient is greater than the suppression effect of rotational speed gradient, and the convection of the small granules occurs naturally. 
	If $a > a_\mathrm{ad}$, the temperature gradients alone are not sufficient to drive convection of the large granules.
	The large granules may generate convection with the assistance of other forms of convection, or break into smaller granules.
	The larger the size, the weaker the convection and the smaller the number of the large granules.
    In summary, for granules in a certain solar quiet region, the conclusions related to convection are as follows:
    \begin{itemize}
   	\item[] (1)  
   	According to the new convection criterion, there is a critical granular size $a_\mathrm{ad}$ as the boundary for dividing granules into large granules with $a > a_\mathrm{ad}$ and small granules with $a < a_\mathrm{ad}$;
    \item[] (2)  
    For the small granules with $a < a_\mathrm{ad}$, due to natural validity of the new convection criterion, they are in natural convection.
    For the large granules with $a > a_\mathrm{ad}$,  due to the invalidity of the new convection criterion and the auxiliary drive of other forms of convection, they are in forced convection.
    \item[] (3)  
    As the suppression effect of the rotation increases with granular size, the number of the large granules in forced convection decreases rapidly with the increase of granular size, while the number of the small granules in natural convection increases monotonously or remains flat with the decrease of granular size. 		
	\end{itemize}
	The observed photospheric granules are considered to be the result of the air mass overshoot at the top of the troposphere, and the above conclusions are consistent with the observed phenomena of photospheric granulation in solar quiet regions. 
	However, some models believe that convection is caused by the ionization and recombination of hydrogen atoms in an area about $2000 \mathrm{m}$ thick below the photosphere layer.
	In this area, the temperature gradient does not reach the adiabatic temperature gradient, but the ionicity gradient replaces the temperature gradient as the energy source for convection.
	Since the ionicity gradient has the same properties as the temperature gradient and is independent of the size of granules, the broad temperature gradient including the ionicity gradient still applies to the above discussions and the results remain unchanged.
	
	\subsection{Solar granule size effected by rotation in the solar active region}
	
	In Case(1.2), it can be seen from Eq.(\ref{Eq.30}) that the rotational speed $\Omega$ and the relative rotational speed gradient $\frac{\mathrm{d} \Omega}{\Omega \mathrm{d} l}$ will effect the critical size $a_\mathrm{ad}$.
	If $\Omega$ and $\frac{\mathrm{d} \Omega}{\Omega \mathrm{d} l}$ increase, then $\Omega^2 \left( \frac{\mathrm{d} \Omega}{\Omega \mathrm{d} l} - \frac{2}{3} \frac{\mathrm{d} \rho}{\rho \mathrm{d} l} \right)$ will increase and further $a_\mathrm{ad}$ will decrease.
	The value of critical size $a_\mathrm{ad}$ represents a limit on the number of the granules participating in thermal convection, so the critical size $a_\mathrm{ad}$ is positively correlated with the Reynolds number $Re$ of thermal convection, that is, the smaller $a_\mathrm{ad}$, the smaller $Re$. 
	Intuitively, in a certain area, if $a_\mathrm{ad}$ decreases, the heat transport efficiency of the granules will be reduced, which means that $\frac{\mathrm{d} T}{T \mathrm{d} l} - \left( \gamma - 1 \right) \frac{\mathrm{d} \rho}{\rho \mathrm{d} l}$ will decrease, then relative temperature gradient $\frac{\mathrm{d} T}{T \mathrm{d} l}$
	and further temperature $T$ will decrease. 
	In short, the larger $\Omega$ and $\frac{\mathrm{d} \Omega}{\Omega \mathrm{d} l}$, the smaller $a_\mathrm{ad}$, $\frac{\mathrm{d} T}{T \mathrm{d} l}$ and $T$.
	
    Sunspots in the solar active region have higher rotational speeds and lower temperatures than in the adjacent regions.
    It is generally believed that sunspots are caused by the solar magnetic field.
    Suppose that in an area on the top of the solar convection zone, the solar magnetic field increases the rotational speed here.
    As $\Omega$ increases, $\frac{\mathrm{d} \Omega}{\Omega \mathrm{d} l}$ increases, then $a_\mathrm{ad}$ and $T$ decreases. 
    Finally, there is a low-temperature area called the sunspot.
	The significant decrease in $a_\mathrm{ad}$ and $T$ can explain why the sizes of granules in the solar active region are smaller than those in the quiet region and the temperatures of sunspots are lower .
		
	\section{Conclusions}
	
	The rotation of granules shows thermal properties similar to the thermal motion of molecules, which effects convection criterion and solar granule size.
	The following conclusions can be drawn from the derivations and analyses:
	\begin{itemize}
		\item[] (1)  
		During the spherically symmetric expansion of the individual granule, the granular rotation generates the rotational additional pressure $\bar{p}_\Omega = k_\Omega \rho \Omega^2 a^2$ and corresponding rotational equivalent temperature $T_\Omega =\frac{M_\mathrm{m}}{R_\mathrm{m}} k_\Omega \Omega^2 a^2$. The larger the granule size $a$, the larger the rotational additional pressure $\bar{p}_\Omega$ and the rotational equivalent temperature $T_\Omega$.
		\item[] (2)  
		Considering the rotational equivalent temperture $T_\Omega$, the new convection criterion $\left| \frac{\mathrm{d} T  \, \text{+} \, \mathrm{d} T_\Omega}{\mathrm{d} l} \right| _{\mathrm{rd}} > \left| \frac{\mathrm{d} T  \, \text{+} \,  \mathrm{d} T_\Omega}{\mathrm{d} l} \right| _{\mathrm{ad}}$ is determined by three factors: the temperature gradient $\frac{\mathrm{d} T}{\mathrm{d} l}$, rotational speed gradient $\frac{\mathrm{d} \Omega}{\mathrm{d} l}$, and granule size $a$. 
		If $\frac{\mathrm{d} \rho}{\rho \mathrm{d} l} < 0$, Case(1.1): when $\frac{\mathrm{d} T}{T \mathrm{d} l} > \left( \gamma - 1 \right) \frac{\mathrm{d} \rho}{\rho \mathrm{d} l}$ and $\frac{\mathrm{d} \Omega}{\Omega \mathrm{d} l} > \frac{2}{3} \frac{\mathrm{d} \rho}{\rho \mathrm{d} l}$, no convection; Case(1.2): when $\frac{\mathrm{d} T}{T \mathrm{d} l} < \left( \gamma - 1 \right) \frac{\mathrm{d} \rho}{\rho \mathrm{d} l}$ and $\frac{\mathrm{d} \Omega}{\Omega \mathrm{d} l} > \frac{2}{3} \frac{\mathrm{d} \rho}{\rho \mathrm{d} l}$, $a < a_\mathrm{ad}$; Case(1.3): when $\frac{\mathrm{d} T}{T \mathrm{d} l} > \left( \gamma - 1 \right) \frac{\mathrm{d} \rho}{\rho \mathrm{d} l}$ and $\frac{\mathrm{d} \Omega}{\Omega \mathrm{d} l} < \frac{2}{3} \frac{\mathrm{d} \rho}{\rho \mathrm{d} l}$, $a < a_\mathrm{ad}$; Case(1.4): when $\frac{\mathrm{d} T}{T \mathrm{d} l} < \left( \gamma - 1 \right) \frac{\mathrm{d} \rho}{\rho \mathrm{d} l}$ and $\frac{\mathrm{d} \Omega}{\Omega \mathrm{d} l} < \frac{2}{3} \frac{\mathrm{d} \rho}{\rho \mathrm{d} l}$, no limit on $a$. Where the critical size $a_\mathrm{ad}$	meets
		\begin{equation}\label{Eq.31}
			a^2_\mathrm{ad} = \frac
			{- T \left[ \frac{\mathrm{d} T}{T \mathrm{d} l} - \left( \gamma - 1 \right) \frac{\mathrm{d} \rho}{\rho \mathrm{d} l} \right]}
			{\frac{2 k_\Omega M_m}{ R_m} \Omega^2 \left( \frac{\mathrm{d} \Omega}{\Omega \mathrm{d} l} - \frac{2}{3} \frac{\mathrm{d} \rho}{\rho \mathrm{d} l} \right)} .
		\end{equation}
		\item[] (3)  
		The convection criterion for granules in most of the solar convection zone belongs to Case(1.2). 
		In the solar quiet region, the larger the size $a$, the weaker the convection and the smaller the number of the granules. 
		Taking the critical size $a_\mathrm{ad}$ as boundary, the small granules with $a < \mathrm{ad}$ are in natural convection, and the large granules with $a > \mathrm{ad}$ are in forced convection.
		In the solar active region, the larger the rotational speed $\Omega$, the smaller the critical size $a_\mathrm{ad}$ and temperature $T$. 
		The solar magnetic field increases the rotational speed , making sunspots have smaller granule size and lower temperature.		
	\end{itemize}
		
	The hypothesis of rotational equivalent temperature can well explain some phenomena in the solar convection zone, and may also be applied to boundary of the solar convection zone. 
	In other words, this theory may help us to further explore the causes of solar differential rotation and sunspots.

	
	%
	
	\section*{Data Availability}
	The data underlying this article are available in the article and in its online supplementary material.
	%
	%

	
	
	\bibliographystyle{mnras}
	\bibliography{referencegranule} 

	
	
	
	
	%
	

	\bsp	
	\label{lastpage}
\end{document}